\documentclass[conference]{IEEEtran}
\usepackage{amsmath,amssymb}
\usepackage{graphicx}
\usepackage{booktabs}
\usepackage{balance}
\usepackage{cite}
\usepackage{xurl}
\usepackage[hidelinks]{hyperref}

\newtheorem{proposition}{Proposition}

\title{Adaptive Budget Allocation for Smart-Contract Fuzzing:
Why Online Feedback Fails to Improve Vulnerability Triage}
\author{%
  \IEEEauthorblockN{\makebox[\columnwidth]{Yanzi Liu}}
  \IEEEauthorblockA{%
    \textit{School of Biomedical Engineering \& Imaging Sciences}\\
    \textit{King's College London}\\
    London, UK\quad\href{mailto:yanzi.liu@kcl.ac.uk}{yanzi.liu@kcl.ac.uk}%
  }
  \and
  \IEEEauthorblockN{\makebox[\columnwidth]{Jiachi Tong}}
  \IEEEauthorblockA{%
    \textit{Department of Mathematics}\\
    \textit{Imperial College London}\\
    London, UK\quad\href{mailto:jt2522@ic.ac.uk}{jt2522@ic.ac.uk}%
  }%
}

\AddToHookNext{shipout/foreground}{%
  \put(\dimexpr1in+\hoffset+\oddsidemargin\relax,
       -\dimexpr\paperheight-0.70in\relax){%
    \parbox[t]{\textwidth}{%
      \normalfont\fontsize{8}{9}\selectfont\raggedright
      \hyphenpenalty=10000\exhyphenpenalty=10000
      \textcopyright{} 2026 IEEE. Personal use of this material is permitted. Permission from IEEE must be obtained for all other uses, in any current or future media, including reprinting/republishing this material for advertising or promotional purposes, creating new collective works, for resale or redistribution to servers or lists, or reuse of any copyrighted component of this work in other works.\par
    }%
  }%
}

\begin{document}
\maketitle

\begin{abstract}
Smart-contract fuzzing campaigns share a finite budget and vary widely in the execution cost of a first correct-class vulnerability report. Using trace replay on 143 annotated and 350 bug-injected contracts, we test whether online feedback improves triage. At a mean budget of 100 executions per contract, uniform allocation reaches 72\% and 70\% of the respective clairvoyant optima. All eight online policies trail uniform on aggregate triage, as do the seven repeated on the bug-injected corpus. Five online signals yield ranking AUCs of 0.32--0.51 for imminent reports. Synthetic-oracle experiments show that sufficiently accurate detection-cost estimates recover more of the gap than detectability rankings. This motivates an offline probability-curve allocator that predicts first-report probabilities across budgets and allocates by predicted gain per execution. In a frozen evaluation on four fresh seeds per corpus, detection--budget AUC improves over uniform by 5.38\% and 4.81\% on the two source corpora and by 1.99--6.54\% on three additional public benchmark subsets.
\end{abstract}


\section{Introduction}
\label{sec:intro}

Smart contracts can hold assets and execute irreversible state transitions, making pre-deployment analysis important. ContractFuzzer~\cite{contractfuzzer}, Harvey~\cite{harvey}, ItyFuzz~\cite{ityfuzz} and EF/CF~\cite{efcf} improve how effort is spent within a fuzzing campaign. We ask how to divide a shared execution budget across campaigns to maximize the number of contracts with a first correct-class vulnerability report.

Large-scale smart-contract evaluations typically assign the same timeout to every contract~\cite{durieux,smartbugs,solidifi}. In our corpus, the median campaign reaches 95\% of its recorded coverage after 200 executions, while the cost of a first correct-class report spans three orders of magnitude among campaigns that report. This variation creates an opportunity for reallocation if the scheduler can estimate what additional executions will yield.

Adaptive scheduling in general-purpose fuzzing~\cite{woo,aflfast,ecofuzz} motivates testing whether coverage growth, discovery rate, missing mass and detector activity can guide reallocation. Replay reveals a substantial gap between uniform and clairvoyant allocation. Some tested online policies improve coverage, yet all trail uniform on aggregate triage. The measured signals provide little information about imminent correct-class reports.

Synthetic-oracle experiments identify what the measured signals lack. A detectability ranking orders campaigns. A detection-cost estimate also informs how many executions to commit. We extend the supervised static ranker to estimate report-arrival probabilities across budgets. The resulting Offline Probability-Curve Allocator uses prior campaigns to compare the predicted return on additional executions under the same triage objective.

\begin{itemize}
    \item We formulate portfolio-level budget allocation for smart-contract fuzzing and derive a clairvoyant optimum for contract-level triage, computable in \(O(N\log N)\) time by sorting detection costs.
    \item Our Ethereum Virtual Machine (EVM) fuzzer produces the same \(n_c\)-execution trace prefix across slice partitions, enabling controlled policy comparisons by replay.
    \item We test eight online policies on 143 SmartBugs-curated contracts and repeat seven on 350 SolidiFI contracts. None improves aggregate triage over uniform, although coverage improves by up to 14.4\% at 100 executions per contract. Signal analysis and synthetic oracles identify the value of budget-scale predictions.
    \item We train an offline allocator on the two source corpora to allocate by predicted probability gain per execution. Frozen evaluation on four fresh seeds shows gains on both source corpora and three additional subsets (Sec.~\ref{sec:offline-eval}).
\end{itemize}

\section{Background and Related Work}

\subsection{Smart-contract fuzzing}

ContractFuzzer~\cite{contractfuzzer} established ABI-driven dynamic testing with vulnerability oracles. Other approaches use property-based testing~\cite{echidna}, branch-distance guidance~\cite{sfuzz} and data-flow guidance~\cite{confuzzius}. Harvey~\cite{harvey} improves transaction-sequence exploration, ItyFuzz~\cite{ityfuzz} uses state snapshots, and EF/CF~\cite{efcf} improves execution throughput. These techniques guide testing within individual campaigns. Echidna-parade~\cite{parade} diversifies configurations across cores for a single target. Our portfolio scheduler divides a shared budget among campaigns for different contracts.

\subsection{Adaptive scheduling and progress signals}

Woo et al.~\cite{woo} formulate mutation scheduling as a multi-armed bandit, AFLFast~\cite{aflfast} allocates energy according to path-transition frequency, and EcoFuzz~\cite{ecofuzz} uses adversarial bandits. These approaches rely on online proxies for future returns. We test whether such proxies are informative when the scheduled unit is a contract campaign and the return is a first correct-class report.

Coverage-based progress estimators provide candidate proxies. STADS~\cite{stads} treats fuzzing as species discovery and uses missing-mass estimators to quantify unseen behavior. Residual-risk estimation~\cite{residual} estimates how much behavior remains undiscovered. For portfolio triage, the relevant empirical question is whether these measures predict when a correct-class report will appear.

\subsection{Benchmarks and evaluation protocol}

We evaluate on SmartBugs~\cite{smartbugs}, whose curated set has supported comparative tool studies~\cite{durieux}, and SolidiFI~\cite{solidifi}. Both provide labeled vulnerable contracts suitable for reproducible local execution. Uniform allocation follows the conventional fixed per-contract budget and serves as the reference under the same total budget.

\section{Problem Formulation}
\label{sec:problem}

Consider a portfolio of $N$ contracts $\mathcal{C}=\{1,\ldots,N\}$. One unit of
budget is one transaction execution. Campaign $c$ can be observed up to a
recorded horizon $L_c$. An allocation is a vector $\mathbf{n}=(n_1,\ldots,n_N)$
such that $0\leq n_c\leq L_c$ and $\sum_c n_c\leq B$.

Let $D_c\in\{1,\ldots,L_c\}\cup\{\infty\}$ denote the first execution at which
campaign $c$ produces a report whose class matches the contract's ground-truth
annotation. If no such report appears within $L_c$, then $D_c=\infty$.
Following the contract-level reporting used in large-scale smart-contract tool
evaluations~\cite{smartbugs,solidifi}, we measure triage breadth as
\begin{equation}
  \mathrm{Det}(\mathbf{n})=\bigl|\{c:n_c\geq D_c\}\bigr|,
  \label{eq:objective}
\end{equation}
the number of contracts with at least one correct-class report. An online
policy repeatedly selects a campaign, allocates a slice of $\Delta$ executions
and observes label-free feedback from that campaign. We use ground-truth
annotations and $D_c$ to score the allocation after replay. During allocation,
both the online policies and the offline predictors lack access to these
outcomes for the evaluated campaign.

\begin{proposition}[Clairvoyant optimum]
Under the budget and horizon constraints above, the maximum of
\eqref{eq:objective} is the largest $k$ for which the $k$ smallest finite
values of $D_c$ sum to at most $B$.
\end{proposition}

\begin{IEEEproof}[Proof sketch]
A detected contract contributes one unit to \eqref{eq:objective} once it
receives at least $D_c$ executions. Allocating more than $D_c$ does not increase
the objective, while allocating less contributes nothing. An optimum therefore
funds a subset $S$ with exactly $D_c$ executions per selected contract and
zero elsewhere, maximizing $|S|$ under $\sum_{c\in S}D_c\leq B$. Since every
selected contract has the same value, the cheapest-first ordering is optimal.
Sorting the finite costs requires $O(N\log N)$ time.
\end{IEEEproof}

The bound measures the headroom available to any allocator on the recorded
traces and is clairvoyant because it observes every $D_c$. It is exact for the
replay benchmark and a lower bound on the hindsight optimum for uncensored
campaigns when reports would occur beyond the recorded horizon. We also report
total coverage and raw detector reports as secondary objectives.

Equation~\eqref{eq:objective} rewards completion. A campaign contributes one
unit at $D_c$, and any additional executions leave its contribution unchanged.
A detectability ranking gives an order of priority. To decide how much budget
to commit, the allocator must also estimate whether a report may require 10
more executions or 1000. Portfolios with the same ranking can therefore require
different allocations under the same budget. This motivates the comparison of
rankings and cost estimates in Sec.~\ref{sec:signals}.

\section{Measurement Infrastructure}
\label{sec:infrastructure}

Comparing allocation policies by replay requires the same campaign trajectory
under different divisions of its execution budget. We describe the fuzzer and
its reporting criterion, then establish the slicing property that enables
replay.

\subsection{Instrumented EVM Fuzzer}

We implement a greybox fuzzer on a Python EVM configured at the Byzantium
fork. Transactions are applied as raw messages with a gas limit of
$2\times 10^6$, removing signature and nonce handling from the hot loop without
changing contract semantics. Each campaign deploys every deployable contract in
a benchmark file, funds three externally owned accounts and three harness
contracts, then repeatedly resets to the post-deployment state and executes a
mutated transaction sequence. Sequences that cover new branches are retained
for further mutation. Arguments are generated according to ABI types, and
harness addresses are included in the address pool so that callback-based
behavior remains reachable.

The three harness contracts exercise bounded re-entrancy, failure of low-level
calls, and passive acceptance. Execution takes place in a local, isolated EVM.

\subsection{Vulnerability Oracles}

The interpreter exposes eight dynamic detectors: unchecked low-level calls,
re-entrancy, integer overflow or underflow reaching persistent storage,
unprotected \texttt{SELFDESTRUCT}, ether leak, unprotected ownership change,
assertion failure and block-state dependence. The differential block-state
detector compares balances and storage writes when replaying a sequence from
the same snapshot with changes to timestamp, block number, difficulty and
coinbase. Replay executions count toward the campaign budget.

A contract is counted as detected when a report matches its annotated class.
Because this criterion evaluates class agreement at the contract level, a
same-class false positive can be credited. We refer to such class-matched
outputs as correct-class reports. This criterion is coarser than the original
SolidiFI protocol~\cite{solidifi} and provides a consistent contract-level
objective across both benchmarks.

We calibrate the detector suite against the six published tools whose raw
SolidiFI outputs are distributed with the benchmark: Oyente, Securify,
Mythril, SmartCheck, Manticore and Slither. Re-scoring those reports under our
contract-level criterion shows that our suite matches or exceeds the published
tool in 12 of 21 tool--class comparisons across the five classes with
dedicated detectors (Table~\ref{tab:solidifi-calibration}). This comparison
establishes the suite's detection capability under our criterion. SolidiFI
contracts contain many injected bugs of one class, and the static tools are
not budget-limited. We report the \texttt{tx.origin} row for completeness and
exclude it from the head-to-head count because our suite has no dedicated
detector for that class.

At the recorded horizon, the suite produces a correct-class report on
53.8\% of SmartBugs-curated contracts and 79.1\% of SolidiFI contracts. Four
curated classes are never detected, and unchecked low-level calls contribute
a large share of reports, motivating the per-class analysis in
Sec.~\ref{sec:evaluation}.

\begin{table}[t]
\caption{Detection profile of the oracle suite on SmartBugs-curated at the
6000-execution horizon, averaged over eight seeds. A contract is credited only
when the reported class matches its annotation.}
\label{tab:smartbugs-profile}
\centering
\small
\setlength{\tabcolsep}{4pt}
\renewcommand{\arraystretch}{1.06}
\begin{tabular*}{\columnwidth}{@{\extracolsep{\fill}}lrrr@{}}
\toprule
Vulnerability class & Contracts & Detected & Rate \\
\midrule
access control            & 18 & 10.5 & 58\% \\
arithmetic                & 15 & 11.2 & 75\% \\
bad randomness            &  8 &  1.0 & 12\% \\
denial of service         &  6 &  0.0 &  0\% \\
front running             &  4 &  0.0 &  0\% \\
other                     &  3 &  0.0 &  0\% \\
reentrancy                & 31 & 11.5 & 37\% \\
short addresses           &  1 &  0.0 &  0\% \\
time manipulation         &  5 &  1.0 & 20\% \\
unchecked low level calls & 52 & 41.6 & 80\% \\
\midrule
All                       & 143 & 76.9 & 54\% \\
\bottomrule
\end{tabular*}
\end{table}

\subsection{Slicing Invariance and Replay}

Each campaign owns an independent EVM state, corpus and pseudo-random stream.
Its next action depends on the campaign history and is independent of the
scheduler's slice size. Consequently, after consuming $n$ executions, the
campaign is in the same state whether those executions were allocated
continuously or across multiple slices. We verify this property by comparing
\texttt{step(1500)}, $30\times\texttt{step(50)}$ and the irregular sequence
$(7,113,400,980)$. All three produce identical per-execution coverage,
missing-mass and report logs.

The differential block-state oracle introduces one bounded exception.
Replaying a transaction sequence can require up to $3L_{\mathrm{seq}}$
additional executions, where $L_{\mathrm{seq}}\leq8$, before control returns
to the scheduler. The associated report is timestamped at the beginning of the
replay batch. Replay is therefore exact at recorded-execution granularity and
approximate at slice boundaries. We quantify this effect by repeating the main
budget sweep with the block-state detectors removed from both scoring and the
report-feedback stream.

Each SmartBugs campaign is recorded once for 6000 executions and each SolidiFI
campaign for 4000 executions. At every execution we store newly covered
species, the singleton and doubleton counts $f_1,f_2$ used by missing-mass
estimators, cumulative coverage, an oracle-aligned species space used by
Sec.~\ref{sec:policies}, elapsed wall-clock time and distinct report indices.
A per-campaign wall-clock cap truncates 16 of the $143\times8$ curated traces,
with the shortest trace containing 450 executions. Policies respect the
recorded horizon. The resulting corpus contains 6.8M curated and 5.6M SolidiFI
executions. Once recorded, additional policy and budget evaluations require
only trace replay.

Every policy purchases prefixes of the same campaign trajectories. Campaign
state advances only when funded, preserving mutation streams across
interruptions and removing a major source of between-policy variance. The same
traces support repeated sweeps across seeds, budgets and sensitivity settings.

\subsection{Benchmarks}

SmartBugs-curated~\cite{smartbugs} contains 143 contracts labeled by DASP
vulnerability class. SolidiFI~\cite{solidifi} injects one vulnerability class
into each of 350 contract copies across seven classes. SmartBugs contracts use
the recorded compiler versions. SolidiFI's loose pragmas are normalized to
Solidity 0.5.17.

\begin{figure*}[!t]
  \centering
  \includegraphics[width=\textwidth]{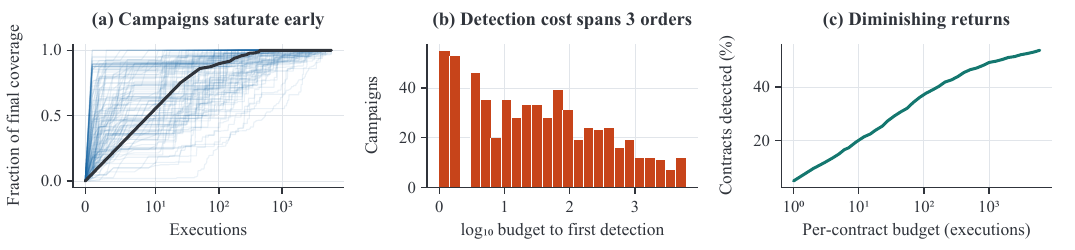}
  \caption{Portfolio heterogeneity motivates budget reallocation. (a) Campaigns approach their recorded coverage at different rates. (b) The execution cost of a first correct-class report varies widely among campaigns that report. (c) Triage improves rapidly at small budgets and then exhibits diminishing returns. These patterns establish the opportunity for reallocation.}
  \label{fig:headroom}
\end{figure*}

\begin{table}[!t]
  \centering
  \caption{External calibration on SolidiFI. Values are the percentages of contracts for which each tool emits at least one report of the injected class, re-scored under our contract-level criterion. A dash indicates a tool--class pair without a result in the published artifact~\cite{solidifi}.}
  \label{tab:solidifi-calibration}
  \small
  \setlength{\tabcolsep}{2pt}
  \renewcommand{\arraystretch}{1.06}
  \begin{tabular*}{\columnwidth}{@{\extracolsep{\fill}}lrrrrrrr@{}}
    \toprule
    Injected class & \textbf{Ours} & Oyente & Sec. & Mythril & SChk. & Mant. & Slither \\
    \midrule
    reentrancy     & \textbf{86} & 100 & 92 & 76 & 0 & 46 & 100 \\
    unhandled exc. & \textbf{98} & 100 & 90 & 78 & 52 & -- & 100 \\
    unchecked send & \textbf{88} & -- & 92 & 82 & -- & -- & -- \\
    overflow       & \textbf{97} & 100 & -- & 76 & 96 & 68 & -- \\
    timestamp      & \textbf{88} & 0 & -- & 78 & 94 & -- & 100 \\
    tx.origin      & \textbf{96} & -- & -- & 78 & 68 & -- & 100 \\
    TOD            & \textbf{0} & 100 & 92 & -- & -- & -- & -- \\
    \bottomrule
  \end{tabular*}
\end{table}

\section{Allocation Policies}\label{sec:policies}

The online policies use label-free execution feedback to update which campaign receives the next slice. The supervised policies learn from labeled historical outcomes and allocate budget from static contract information. For campaign $c$, let $n_c$ denote its executions consumed, $S_c$ its covered species count, $f_{1,c}$ the number of species observed exactly once, and $F_c$ the cumulative number of raw detector reports of any class.

\textbf{Uniform and random.}
Uniform allocation assigns the same execution budget to every campaign and serves as the reference protocol. Random allocation selects a campaign uniformly at random for each slice.

\textbf{Coverage-driven policies.}
Discovery-rate greedy and $\epsilon$-greedy prioritize the number of newly covered species per execution over a trailing window. UCB1 and EXP3 treat campaigns as bandit arms and use the same coverage reward, following established scheduling formulations in fuzzing~\cite{woo,ecofuzz}.

\textbf{Good--Turing marginal yield.}
This policy replaces the empirical discovery rate with an add-one-smoothed missing-mass estimate $\hat{p}_c=(f_{1,c}+1)/(n_c+1)$~\cite{stads}. It multiplies this estimate by an empirical-Bayes estimate of reports per newly covered species, $\hat{y}_c=(F_c+\alpha\mu)/(S_c+\alpha)$, where $\mu$ is the portfolio mean. The policy downweights campaigns that have produced a report by $\delta=0.05$ and allocates budget proportionally to the resulting score. Proportional allocation avoids repeatedly funding only the highest-scoring campaign when Eq.~\eqref{eq:objective} rewards completing campaigns.

\textbf{Satisficing.}
Satisficing shares the budget fairly among campaigns that have yet to produce any report.

\textbf{Oracle-aligned missing mass.}
This policy retains the missing-mass rule and replaces branch coverage with events used by the vulnerability oracles. These include external-call sites and outcomes, value-carrying calls, written storage slots, arithmetic-overflow sites, self-destruct sites, re-entrant frames, block-state reads, and revert kinds. We also test the sign-inverted variant motivated by Sec.~\ref{sec:signals}.

\textbf{Supervised static ranker.}
The ranker predicts the probability that a contract produces a first correct-class report within the per-contract budget. It uses static bytecode features including security-relevant opcode frequencies, code size, and counts of external and payable functions. Training excludes the evaluated seed and uses leave-one-contract-out prediction, so a contract's own outcomes do not contribute to its score.

\textbf{Offline Probability-Curve Allocator.}
The synthetic-oracle comparison in Sec.~\ref{sec:signals} motivates estimating report arrival on an execution scale. The static ranker supplies a score at the reference budget. We extend it to estimate the cumulative probability of a first correct-class report at several execution horizons, allowing the allocator to compare the predicted returns of candidate budget increments.

Two ridge models use opcode and opcode-group features and fuzzer-aware static features, respectively. The latter combine the original features with ABI argument structure, \texttt{PUSH} constants, and control-flow summaries. Both models use portfolio-wise empirical-rank normalization and a penalty of $\lambda=100$. We pool SmartBugs and SolidiFI, weighting each source campaign by $0.7M/143$ and $0.3M/350$, respectively, where $M=493$. At each of the 16 SmartBugs budget levels $h_j$, both models estimate $\Pr(D_c\le h_j)$ using the corresponding frequencies observed across training seeds as targets. Let $r^o_{cj}$ and $r^f_{cj}$ denote the two raw predictions. We combine them into monotone curve nodes
\begin{equation}
  q_c(h_j)=\left[\max_{\ell\le j}
  \operatorname{clip}_{[0,1]}\!\left(0.6r^o_{c\ell}+0.4r^f_{c\ell}\right)\right]^{1.25}.
  \label{eq:curve}
\end{equation}
We set $q_c(0)=0$ and interpolate linearly between nodes.

Starting from $\lfloor 0.15B/N\rfloor$ executions per campaign, the allocator repeatedly selects the campaign and future curve endpoint $t$ with the largest predicted gain per execution, $(q_c(t)-q_c(n_c))/(t-n_c)$. It funds that segment or the remaining budget. This phase prefers allocations up to $\lceil 3B/N\rceil$ per campaign. Once positive gains or these preferred limits are exhausted, equal-count round-robin allocates the remaining budget. All phases respect recorded trace capacities. The fixed rule uses the same parameters for every dataset and budget and allocates using static features, available capacities, and $B$. The curves remain fixed during execution. Section~\ref{sec:offline-eval} describes the training exclusions and evaluation.

\section{Evaluation}\label{sec:evaluation}

\subsection{Experimental setup}
We use the same replay objective to quantify the headroom over uniform allocation, diagnose the tested online policies, and evaluate a learned report-arrival curve.

We record eight independent SmartBugs campaigns per contract and four SolidiFI campaigns per contract, all CPU-only. The SmartBugs sweep uses 16 budget levels from 5 to 2000 executions, and SolidiFI uses 10 levels from 5 to 800. Budgets are reported as $k=B/N$, the per-contract share under uniform allocation. Adaptive policies use slices of $\Delta=5$ executions, with $\Delta\in\{1,25\}$ for sensitivity analysis. We average results over seeds and report bootstrap 95\% confidence intervals. To account for dependence across budgets in the original sweep, we compute each seed's detection--budget AUC on a logarithmic budget axis and compare it with uniform allocation using a paired Wilcoxon signed-rank test. We also report absolute triage breadth, the fraction of the clairvoyant optimum reached by uniform allocation, and coverage as a control.

\begin{figure}[!t]
  \centering
  \includegraphics[width=\columnwidth]{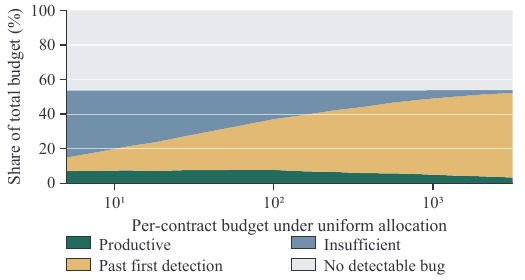}
  \caption{Uniform allocation spends an increasing fraction of the budget where it cannot create an additional triaged contract. Productive executions move a campaign toward its first correct-class report. The remaining budget is spent after reporting, below the required detection cost, or on campaigns that do not report within the recorded horizon.}
  \label{fig:spending}
\end{figure}

\subsection{RQ1: How much headroom does uniform allocation leave?}
On SmartBugs, campaigns saturate at different rates and the marginal return from additional budget decreases sharply (Fig.~\ref{fig:headroom}). Among campaigns that report, the 25th, 50th, 75th, and 95th percentile detection costs are 4.5, 27.0, 170.5, and 1939.0 executions, respectively. Of all campaigns, 46\% do not report within their recorded horizon.

On SmartBugs, the gap to the clairvoyant optimum persists across the budget range. At $k=20$, uniform allocation triages 35.2 contracts against the optimum's 62.1, or 57\% of the trace-level bound. At $k=100$, the corresponding values are 53.2 and 73.8 (72\%). At $k=400$, they are 64.8 and 76.9 (84\%). On SolidiFI, uniform allocation reaches 70\% of the optimum at $k=100$. The largest opportunity for reallocation occurs at scarce budgets, when online policies also have the least time to collect campaign-specific feedback.

Fig.~\ref{fig:spending} shows where the budget goes. At larger $k$, a growing fraction goes to campaigns that have already reported or do not report within the horizon. At smaller budgets, many campaigns stop before their detection cost. The clairvoyant optimum funds the campaigns with the lowest completion costs. A practical allocator needs information about report arrival to identify which budget increments are likely to yield an additional triaged contract.

\begin{table}[!t]
  \centering
  \caption{Contracts triaged on SmartBugs-curated at five per-contract budgets, averaged over eight seeds. $\ast$ denotes $p<0.05$ versus uniform allocation under an uncorrected paired Wilcoxon test. Every starred adaptive result is lower than uniform allocation.}
  \label{tab:online-policies}
  \footnotesize
  \setlength{\tabcolsep}{2pt}
  \renewcommand{\arraystretch}{1.04}
  \newcommand{\triaged}[2]{#1\makebox[0.65em][l]{\textsuperscript{#2}}}
  \begin{tabular*}{\columnwidth}{@{\extracolsep{\fill}}lrrrrr@{}}
    \toprule
    Allocation policy & $k=20$ & $k=50$ & $k=100$ & $k=200$ & $k=400$ \\
    \midrule
    Uniform (standard practice) & \triaged{35.2}{} & \triaged{45.1}{} & \triaged{53.2}{} & \triaged{59.2}{} & \triaged{64.8}{} \\
    Random & \triaged{32.9}{*} & \triaged{45.0}{} & \triaged{53.2}{} & \triaged{58.8}{} & \triaged{65.0}{} \\
    Discovery-rate greedy & \triaged{33.2}{*} & \triaged{43.5}{*} & \triaged{43.5}{*} & \triaged{45.1}{*} & \triaged{46.8}{*} \\
    $\epsilon$-greedy & \triaged{33.1}{*} & \triaged{44.8}{} & \triaged{46.5}{*} & \triaged{49.8}{*} & \triaged{54.6}{*} \\
    UCB1 & \triaged{32.9}{*} & \triaged{44.1}{} & \triaged{52.9}{} & \triaged{58.5}{} & \triaged{64.5}{} \\
    EXP3 & \triaged{34.1}{} & \triaged{44.1}{} & \triaged{53.1}{} & \triaged{59.5}{} & \triaged{64.8}{} \\
    Satisficing & \triaged{34.9}{} & \triaged{42.8}{*} & \triaged{45.6}{*} & \triaged{46.9}{*} & \triaged{47.4}{*} \\
    Good--Turing marginal yield & \triaged{34.0}{} & \triaged{45.1}{} & \triaged{51.5}{*} & \triaged{57.8}{*} & \triaged{62.8}{*} \\
    Oracle-aligned missing mass & \triaged{33.1}{*} & \triaged{44.0}{} & \triaged{49.2}{*} & \triaged{56.1}{*} & \triaged{61.4}{*} \\
    \quad same signal, inverted & \triaged{24.0}{*} & \triaged{26.4}{*} & \triaged{26.5}{*} & \triaged{26.8}{*} & \triaged{29.0}{*} \\
    Supervised static ranker & \triaged{35.1}{} & \triaged{43.2}{} & \triaged{51.8}{} & \triaged{57.2}{*} & \triaged{61.9}{*} \\
    \midrule
    Clairvoyant optimum & \triaged{62.1}{} & \triaged{69.5}{} & \triaged{73.8}{} & \triaged{76.5}{} & \triaged{76.9}{} \\
    \bottomrule
  \end{tabular*}
\end{table}

\begin{figure*}[!t]
  \begin{minipage}[t]{\columnwidth}
    \centering
    \includegraphics[width=\linewidth]{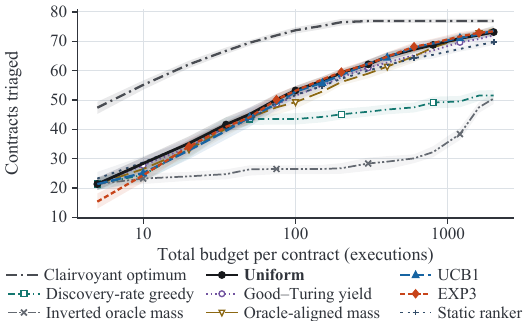}
    \caption{The tested online adaptive policies leave most of the clairvoyant gap unexploited. Across the SmartBugs-curated budget range, their triage curves overlap with or fall below uniform allocation, while the clairvoyant optimum remains substantially higher. Curves show means over eight seeds with bootstrap 95\% confidence intervals.}
    \label{fig:triage}
  \end{minipage}\hfill
  \begin{minipage}[t]{\columnwidth}
    \centering
    \includegraphics[width=\linewidth]{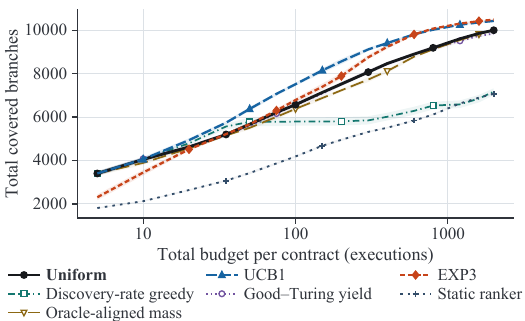}
    \caption{The same allocation policies produce different outcomes when evaluated on coverage. UCB1 and EXP3 improve over uniform allocation on covered block-and-branch species, demonstrating that the scheduler can exploit exploration feedback.}
    \label{fig:coverage}
  \end{minipage}
\end{figure*}

\subsection{RQ2: Do the tested online policies exploit the headroom?}
On SmartBugs, every tested online adaptive policy scores below uniform allocation on the aggregate per-seed statistic. EXP3 is best at $-1.7$\%, with a paired 95\% CI of $[-1.15,-0.48]$ contracts ($p=0.016$). Inverted oracle-aligned missing mass is worst at $-44.2$\%. Table~\ref{tab:online-policies} and Fig.~\ref{fig:triage} show the pattern across budgets. Among 144 comparisons in the original per-budget sweep, 91 favor uniform allocation at nominal significance and two favor an alternative. Under a global null at $\alpha=0.05$, about 3.6 significant comparisons are expected in each direction without correction, so the isolated gains provide little evidence of a systematic improvement.

The two nominally significant gains occur in different regimes. EXP3 improves at $k=1600$, when campaigns are close to exhaustion and the absolute gain is below one contract. The supervised static ranker improves at $k=5$, the only positive result that also appears on SolidiFI. It gains 8.8\% on SmartBugs-curated ($p=0.039$) and 14.7\% on SolidiFI. At this budget, a static prior is useful before a campaign can accumulate much feedback. By $k=100$, the ranker is $-1.9$\% relative to uniform allocation under cross-validation ($p=0.271$, ranking AUC 0.78).

The online result is stable across the main robustness checks. Removing block-state detectors from scoring and feedback leaves uniform allocation at 72\% of the optimum at $k=100$, while the best of seven adaptive policies remains within $[-2.9,0.2]$\% of uniform allocation across five budgets. With $\Delta=1$ and $\Delta=25$, the best of six policies gains 0.9\% and 0.2\%, respectively, at $k=100$. Across sub-portfolios of 20, 50, and 100 contracts and $k\in\{20,50,100\}$, an adaptive policy exceeds uniform allocation in four of 36 cells, always by less than one contract. On SolidiFI, the best adaptive policy is $-1.3$\% relative to uniform allocation at $k=100$.

The conclusion for the tested online policies also holds when the resource unit is wall-clock time. Equal-time allocation, which most closely matches a fixed per-contract timeout, triages 67.2 contracts at 5.0\% of the corpus cost compared with 56.5 under equal executions at the same cost. Of the four adaptive policies replayed in this regime on seeds 1--4, the best is $-13.8$\% relative to equal-time allocation. The equal-time baseline is therefore stronger than the equal-execution baseline in this comparison.

\subsection{RQ3: Does the same feedback improve coverage?}
When the same policies are scored on combined block-and-branch coverage, UCB1 improves by 14.4\% over uniform allocation at $k=100$ ($p=0.0078$), and EXP3 improves at larger budgets (Fig.~\ref{fig:coverage}). The best coverage policy gains 6.8\% at $k=20$, 14.4\% at $k=100$, and 11.2\% at $k=400$. Discovery-rate and $\epsilon$-greedy also improve coverage at small budgets before falling behind at larger budgets. Satisficing and the supervised static ranker do not improve coverage.

The coverage gains show that slice-based scheduling responds to productive exploration. The lack of corresponding triage gains points to a mismatch between this feedback and the report-based objective. We next test whether the measured signals predict imminent reports directly.

\subsection{RQ4: What information about report arrival is useful?}\label{sec:signals}
Table~\ref{tab:signals} measures five candidate signals after a 100-execution prefix. We test whether each signal ranks campaigns that will produce a first correct-class report within the next 50 or 200 executions ahead of campaigns that will not. The tie-aware ranking AUCs range from 0.32 to 0.51. Three signals perform approximately at chance, and the two missing-mass signals tend to rank campaigns in the unhelpful direction. Their positive correlations with the remaining detection cost $D_c-n$ indicate that campaigns with higher missing mass tend to need more executions before reporting.

\begin{table}[!t]
  \centering
  \caption{Predictive value of online allocation signals after 100 executions. $\mathrm{AUC}_h$ measures whether a signal ranks campaigns that have yet to report by whether they report within the next $h$ executions. An AUC of 0.5 indicates chance. Values are averaged over seeds with at least three eligible campaigns (six of eight at $h=50$ and all eight at $h=200$). $\rho$ is the Spearman correlation with remaining detection cost $D_c-n$. A negative correlation would have the useful direction.}
  \label{tab:signals}
  \small
  \setlength{\tabcolsep}{4pt}
  \renewcommand{\arraystretch}{1.06}
  \begin{tabular*}{\columnwidth}{@{\extracolsep{\fill}}lrrr@{}}
    \toprule
    Allocation signal at $n=100$ & $\mathrm{AUC}_{50}$ & $\mathrm{AUC}_{200}$ & $\rho$ \\
    \midrule
    Good--Turing missing mass & 0.41 & 0.51 & $+0.24$ \\
    Oracle-aligned missing mass & 0.32 & 0.41 & $+0.42$ \\
    New species, last 25 executions & 0.48 & 0.51 & $+0.22$ \\
    Species covered so far & 0.44 & 0.49 & $+0.26$ \\
    Raw reports so far & 0.43 & 0.48 & $+0.02$ \\
    \bottomrule
  \end{tabular*}
\end{table}

We test whether reversing the direction improves triage. Allocating budget in inverse proportion to oracle-aligned missing mass yields $-50.2$\% relative to uniform allocation at $k=100$ ($p=0.008$) and $-46.2$\% relative to the forward variant. Inverting the signal deprioritizes campaigns that continue to discover new behavior, and the residual ranking information is too weak to compensate. Together, these measurements show that the tested signals provide weak estimates of imminent report arrival.

\begin{figure}[!t]
  \centering
  \includegraphics[width=\columnwidth]{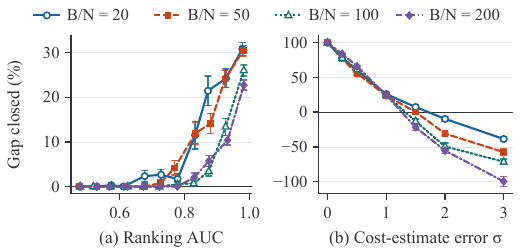}
  \caption{Synthetic oracles compare two kinds of information about report arrival. A perfect ranking recovers part of the uniform-to-clairvoyant gap. Accurate cardinal estimates of detection cost recover a larger share.}
  \label{fig:synthetic}
\end{figure}

We use synthetic oracles to test how much useful information an allocator would gain from knowing campaign ranks or detection costs (Fig.~\ref{fig:synthetic}). With a perfect ranking by detectability and the best of three rank-to-budget mappings (top-$m$ equal sharing, $1/r$, and $1/r^2$), the allocator recovers 29\% of the uniform-to-clairvoyant gap at $k=100$ and 28--33\% across $k\in\{20,50,100,200\}$. At a ranking AUC of approximately 0.9, recovery falls to 10\%. The allocator supplied with a cardinal estimate $\hat{D}_c$ of each campaign's detection cost funds campaigns cheapest first at a fixed safety multiple. It recovers 100\% of the gap with exact costs, 78\% at $\sigma=0.25$, 61\% at $\sigma=0.5$, and 25\% at $\sigma=1$. It becomes harmful at $\sigma=2$. The safety multiple is selected on held-out seeds, excluding the seed being evaluated. The same ordering holds across all swept budgets.

A cardinal cost estimate indicates how much budget a campaign may need and lets the scheduler compare that cost with the available budget. The degradation from $\sigma=0.25$ to $\sigma=1$ shows how recovery depends on preserving meaningful differences in detection costs. Section~\ref{sec:offline-eval} tests whether a cumulative report-arrival curve learned from prior traces provides budget-scale information that improves triage in practice.

\subsection{RQ5: How sensitive are the online results to class mix and detector latency?}
Unchecked low-level calls account for a large fraction of the detections in Table~\ref{tab:smartbugs-profile}. We therefore re-score the online sweep with equal weight on each of the nine vulnerability classes containing at least three contracts. The metric is the mean fraction triaged across those classes. Uniform allocation obtains 22.6\%. No adaptive policy exceeds this result, and the best is $-1.3$\% relative to uniform allocation. At the individual-class level, an adaptive policy exceeds uniform allocation in three of nine classes, by at most 0.75 contracts on average.

We also re-score the same traces under counterfactual oracles that fire up to $8\times$ earlier and, in a separate dimension, extend detection to campaigns that do not currently report. The four policies in this analysis observe only coverage and never observe $D_c$, so re-scoring is exact for those policies. Across all 21 counterfactual and budget settings, uniform allocation reaches 57--97\% of the clairvoyant optimum, and the best adaptive policy ranges from $-4.5$\% to 1.7\% relative to uniform allocation. It exceeds uniform allocation in eight settings, with a maximum gain of 1.7\%. The online result persists across these changes in class mixture and detector latency. The counterfactuals retain the fuzzer's measured coverage trajectories, so the result remains specific to its exploration behavior.

\subsection{Offline evaluation with report-arrival curves}\label{sec:offline-eval}
Following the synthetic-oracle comparison, we test whether learned report-arrival curves improve triage over uniform allocation across execution budgets. The curve models pool the two source corpora. The static ranker trains on each separately. We also evaluate compilable, class-mapped subsets of Not-So-Smart-Contracts (NSSC)~\cite{nssc}, SWC Registry~\cite{swc}, and the ContractFuzzer examples~\cite{contractfuzzer}. These contain 8, 55, and 99 campaigns, respectively, including 28 SWC negative controls. We use the same fuzzer and class-match criterion for all subsets.

\textbf{Training and evaluation.}
Development uses seeds 1--8 on SmartBugs and 1--4 elsewhere. For each source prediction, we exclude the predicted contract's own outcomes and remove the evaluated seed from both training corpora. Predictions for the three additional subsets use coefficients fitted on the two source corpora and unlabeled ranks within the target portfolio. Hyperparameter search prioritizes the smaller AUC gain across the two sources and also uses results from the additional subsets during development.

We freeze the fixed allocator and a budget-specific variant before generating four new seeds (9--12) on all five corpora. Fresh-seed source predictions retain pointwise contract exclusion and use only the old training seeds. The fresh outcomes are used only for evaluation. This yields 2,620 fresh traces on the same contracts. Requested horizons are 6000 executions for SmartBugs and 4000 elsewhere. We retain the original budget grids and use $k\in\{5,10,20,50,100,200,400\}$ for the additional subsets.

\begin{table}[!t]
  \centering
  \caption{Relative detection--budget AUC gain over uniform allocation (\%). Development uses the original seeds and fresh evaluation uses seeds 9--12. Both variants are frozen before the fresh runs. Gains compare mean seed AUCs within each corpus.}
  \label{tab:offline}
  \small
  \setlength{\tabcolsep}{4pt}
  \renewcommand{\arraystretch}{1.06}
  \begin{tabular*}{\columnwidth}{@{\extracolsep{\fill}}lrrr@{}}
    \toprule
    & \multicolumn{2}{c}{Fixed allocator} & Budget-specific \\
    \cmidrule(lr){2-3}\cmidrule(l){4-4}
    Corpus & Development & Fresh & Fresh \\
    \midrule
    SmartBugs (143) & $+5.65$ & \textbf{+5.38} & $+5.15$ \\
    SolidiFI (350) & $+5.89$ & \textbf{+4.81} & $+5.19$ \\
    NSSC (8) & $+5.44$ & \textbf{+1.99} & $-1.26$ \\
    SWC (55) & $+3.84$ & \textbf{+4.20} & $+2.32$ \\
    ContractFuzzer (99) & $+5.03$ & \textbf{+6.54} & $+4.01$ \\
    \bottomrule
  \end{tabular*}
\end{table}

\textbf{Results.}
Table~\ref{tab:offline} reports relative gains based on mean log-budget AUC, denoted by $\bar{A}$ for each allocator. The gain is $100(\bar{A}_{\mathrm{method}}/\bar{A}_{\mathrm{uniform}}-1)$. On the original SmartBugs and SolidiFI traces, the static ranker scores 2.97\% and 2.96\% below uniform allocation, respectively. The curve allocator scores 5.65\% and 5.89\% above it.

On fresh seeds, the fixed allocator gains 5.38\% on SmartBugs and 4.81\% on SolidiFI. The AUC difference is positive for each of the four seeds on both corpora. Gains on the three additional subsets range from 1.99\% to 6.54\%. The fixed allocator exceeds uniform allocation in 19 of 20 corpus--seed comparisons, with one loss on NSSC. The budget-specific variant selects a configuration for each $k$ from development results and applies that configuration across corpora. Its fresh NSSC result is negative, so finer budget-specific selection yields less consistent gains in this evaluation.

The gains reflect the combined changes in training data, features, and allocation. Fresh seeds measure repeatability on the evaluated corpora.

\section{Discussion}
\label{sec:discussion}

Coverage growth and missing mass are weak proxies for imminent reports in our campaigns (Table~\ref{tab:signals}). Uniform per-contract allocation remains a strong baseline against the tested feedback strategies, including in the equal-time comparison. The synthetic-oracle analysis (Sec.~\ref{sec:signals}) suggests estimating how many executions each campaign may need in addition to ranking campaigns by detectability.

The Offline Probability-Curve Allocator estimates report probabilities across execution budgets. The frozen configuration's gains (Table~\ref{tab:offline}) show that these estimates can guide allocation in the evaluated portfolios. Residual-risk estimation~\cite{residual} offers a related approach for coverage. An online extension could initialize report-arrival curves from static features, then update them with execution evidence to reassess the return on further testing.

Measuring discovery across budgets reveals differences that a single timeout can obscure. Replay makes these comparisons inexpensive.

\section{Threats to Validity}
\label{sec:threats}
\balance

\textit{Internal validity.} The study uses one fuzzer and one bespoke oracle suite. We calibrated the suite against six published tools for detection capability. Its report-cost profile reflects our implementation. The coverage control, per-class analysis, and counterfactual oracle sweep test scheduler responsiveness, class mixture, and detector cost. A different exploration strategy could expose a more predictive online signal. Replay is approximate at differential-oracle slice boundaries. The corresponding sensitivity analysis removes these detectors from scoring and feedback. We tested a finite set of online policy configurations, and the negative result concerns those instantiations. The offline search uses development outcomes, followed by a separate evaluation on new seeds with the configurations frozen. Pointwise contract exclusion removes each contract's own training target from its prediction and permits training on other contracts in the same portfolio.

\textit{External validity.} SmartBugs and SolidiFI mainly contain small Solidity~0.4--0.5 contracts. SolidiFI's 350 injected variants derive from 50 base contracts. The additional subsets broaden the corpus mix and share the compilation and class-mapping requirements of the same fuzzer. Campaign observations are limited to their recorded horizons. Larger modern DeFi protocols may have different cost distributions. Our primary budget unit is transaction executions. The equal-time analysis yields qualitatively similar results for the online policies and uses wall-clock times recorded during a parallel trace run. The offline evaluation measures gains in execution budgets, excluding training and feature-extraction costs.

\textit{Construct validity.} Equation~\eqref{eq:objective} measures triage breadth across contracts. Class matching can count same-class false positives because it does not validate source locations. The original aggregate test resolves an approximately uniform $\pm 2\%$ shift in the curve statistic, so the evidence is strongest for effects of that scale. The offline results report observed AUC gains across four fresh seeds per corpus.

\section{Conclusion}
\label{sec:conclusion}

The clairvoyant bound shows room to improve smart-contract triage under a shared execution budget. Eight online policies on SmartBugs and seven repeated on SolidiFI trailed uniform allocation on aggregate triage. The signal and synthetic-oracle experiments point to the value of estimating report arrival on an execution-budget scale.

The Offline Probability-Curve Allocator learns these estimates from prior campaigns and allocates by expected gain per execution. On four fresh seeds per corpus, the frozen allocator improved detection--budget AUC over uniform by 5.38\% on SmartBugs, 4.81\% on SolidiFI, and 1.99--6.54\% on three additional public subsets. These gains support budget-conditioned report prediction for portfolio allocation.

\section*{Artifact Availability}

The fuzzer, original 12.4M-execution corpus, and reproduction scripts will be released under an open-source license. The extended artifact includes the pooled model, features, dataset manifests, frozen configurations, and 2,620 fresh traces.

\bibliographystyle{IEEEtran}
\bibliography{references}

\begin{thebibliography}{10}
\providecommand{\url}[1]{#1}
\csname url@samestyle\endcsname
\providecommand{\newblock}{\relax}
\providecommand{\bibinfo}[2]{#2}
\providecommand{\BIBentrySTDinterwordspacing}{\spaceskip=0pt\relax}
\providecommand{\BIBentryALTinterwordstretchfactor}{4}
\providecommand{\BIBentryALTinterwordspacing}{\spaceskip=\fontdimen2\font plus
\BIBentryALTinterwordstretchfactor\fontdimen3\font minus
  \fontdimen4\font\relax}
\providecommand{\BIBforeignlanguage}[2]{{%
\expandafter\ifx\csname l@#1\endcsname\relax
\typeout{** WARNING: IEEEtran.bst: No hyphenation pattern has been}%
\typeout{** loaded for the language `#1'. Using the pattern for}%
\typeout{** the default language instead.}%
\else
\language=\csname l@#1\endcsname
\fi
#2}}
\providecommand{\BIBdecl}{\relax}
\BIBdecl

\bibitem{contractfuzzer}
B.~Jiang, Y.~Liu, and W.~K. Chan, ``{ContractFuzzer}: Fuzzing smart contracts
  for vulnerability detection,'' in \emph{Proc. ACM/IEEE Int. Conf. Automated
  Software Engineering (ASE)}, 2018.

\bibitem{harvey}
V.~W{\"u}stholz and M.~Christakis, ``{Harvey}: A greybox fuzzer for smart
  contracts,'' in \emph{Proc. ACM ESEC/FSE}, 2020.

\bibitem{ityfuzz}
C.~Shou, S.~Tan, and K.~Sen, ``{ItyFuzz}: Snapshot-based fuzzer for smart
  contract,'' in \emph{Proc. ACM ISSTA}, 2023.

\bibitem{efcf}
M.~Rodler, D.~Paa{\ss}en, W.~Li, L.~Bernhard, T.~Holz, and L.~Davi, ``{EF/CF}:
  High performance smart contract fuzzing for exploit generation,'' in
  \emph{Proc. IEEE EuroS\&P}, 2023.

\bibitem{durieux}
T.~Durieux, J.~F. Ferreira, R.~Abreu, and P.~Cruz, ``Empirical review of
  automated analysis tools on 47,587 {Ethereum} smart contracts,'' in
  \emph{Proc. ICSE}, 2020.

\bibitem{smartbugs}
J.~F. Ferreira, P.~Cruz, T.~Durieux, and R.~Abreu, ``{SmartBugs}: A framework
  to analyze {Solidity} smart contracts,'' in \emph{Proc. IEEE/ACM ASE}, 2020.

\bibitem{solidifi}
A.~Ghaleb and K.~Pattabiraman, ``How effective are smart contract analysis
  tools? evaluating smart contract static analysis tools using bug injection,''
  in \emph{Proc. ACM ISSTA}, 2020.

\bibitem{woo}
M.~Woo, S.~K. Cha, S.~Gottlieb, and D.~Brumley, ``Scheduling black-box
  mutational fuzzing,'' in \emph{Proc. ACM CCS}, 2013.

\bibitem{aflfast}
M.~B{\"o}hme, V.-T. Pham, and A.~Roychoudhury, ``Coverage-based greybox fuzzing
  as {Markov} chain,'' in \emph{Proc. ACM CCS}, 2016.

\bibitem{ecofuzz}
T.~Yue, P.~Wang, Y.~Tang, E.~Wang, B.~Yu, K.~Lu, and X.~Zhou, ``{EcoFuzz}:
  Adaptive energy-saving greybox fuzzing as a variant of the adversarial
  multi-armed bandit,'' in \emph{Proc. USENIX Security}, 2020, pp. 2307--2324.

\bibitem{echidna}
G.~Grieco, W.~Song, A.~Cygan, J.~Feist, and A.~Groce, ``{Echidna}: Effective,
  usable, and fast fuzzing for smart contracts,'' in \emph{Proc. ACM ISSTA},
  2020.

\bibitem{sfuzz}
T.~D. Nguyen, L.~H. Pham, J.~Sun, Y.~Lin, and Q.~T. Minh, ``{sFuzz}: An
  efficient adaptive fuzzer for {Solidity} smart contracts,'' in \emph{Proc.
  ICSE}, 2020.

\bibitem{confuzzius}
C.~F. Torres, A.~K. Iannillo, A.~Gervais, and R.~State, ``{ConFuzzius}: A data
  dependency-aware hybrid fuzzer for smart contracts,'' in \emph{Proc. IEEE
  EuroS\&P}, 2021.

\bibitem{parade}
A.~Groce and G.~Grieco, ``{echidna-parade}: A tool for diverse multicore smart
  contract fuzzing,'' in \emph{Proc. ACM ISSTA}, 2021.

\bibitem{stads}
M.~B{\"o}hme, ``{STADS}: Software testing as species discovery,'' \emph{ACM
  TOSEM}, vol.~27, no.~2, 2018.

\bibitem{residual}
M.~B{\"o}hme, D.~Liyanage, and V.~W{\"u}stholz, ``Estimating residual risk in
  greybox fuzzing,'' in \emph{Proc. ACM ESEC/FSE}, 2021.

\bibitem{nssc}
{Trail of Bits}, ``{Not-so-smart-contracts},''
  \url{https://github.com/crytic/not-so-smart-contracts}, public Solidity
  vulnerability examples, accessed Oct. 2026.

\bibitem{swc}
{Smart Contract Security}, ``{SWC} registry: Smart contract weakness
  classification and test cases,''
  \url{https://github.com/SmartContractSecurity/SWC-registry}, test-case
  snapshot, accessed Oct. 2026.

\end{thebibliography}
\end{document}